\documentclass[10pt]{article}
\usepackage{graphicx}
\usepackage{amsmath}
\usepackage{amssymb}
\usepackage{caption2}
\begin{document}
\bibliographystyle{prsty}
\begin{center}
{\large {\bf \sc{  Analysis of the  vector and axialvector $B_c$ mesons with  QCD sum rules }}} \\[2mm]
Zhi-Gang Wang \footnote{E-mail,zgwang@aliyun.com.  }     \\
 Department of Physics, North China Electric Power University,
Baoding 071003, P. R. China
\end{center}

\begin{abstract}
In this article, we study the vector and axialvector $B_c$ mesons with the QCD sum rules,
and make reasonable predictions for the masses and decay constants, then calculate
the leptonic decay widths. The present predictions for the masses and decay constants can be confronted with the experimental data in the future.
We can also take the masses and decay constants as basic input parameters and study  other phenomenological quantities
with the three-point vacuum correlation functions via the QCD sum rules.
\end{abstract}

 PACS number: 14.40.Pq, 12.38.Lg

Key words: $B_c$ mesons, QCD sum rules

\section{Introduction}

In 1998, the CDF collaboration  observed the pseudoscalar  bottom-charm $B_c$ mesons  through  the decay modes $B_c^{\pm} \to J/\psi \ell^{\pm}X $ and $B_c^{\pm} \to J/\psi \ell^{\pm}\bar{\nu}_{\ell} $ in $p\bar{p}$ collisions at the energy $\sqrt{s}=1.8\,\rm{TeV}$ at the Fermilab Tevatron, the measured mass is $M_{B_c}=(6.40 \pm 0.39 \pm 0.13) \,\rm{GeV}$ \cite{CDF1998}.
In 2007, the  CDF collaboration  observed the pseudoscalar $B_c$ mesons with a significance exceeds  $8\, \sigma$ through the decay modes $B_c^{\pm} \to J/\psi \pi^{\pm}$  in $p\bar{p}$ collisions at the energy  $\sqrt{s}=1.96\,\rm{TeV}$ using the Collider Detector
at Fermilab (CDF II),   the   measured mass is $M_{B_c}=(6275.6 \pm 2.9\pm 2.5) \, \rm{MeV}$ \cite{CDF2008}.  In 2008, the D0 collaboration reconstructed the $B_c^{\pm} \to J/\psi \pi^{\pm}$ decays and observed the pseudoscalar $B_c$ mesons  with a significance more than $5\,\sigma$, the measured  mass is $M_{B_c}=(6300 \pm 14 \pm 5)\,
\rm{ MeV}$ \cite{D02008}. Now the average value listed in the  Review of Particle Physics is  $M_{B_c}=(6.277 \pm 0.006)\,\rm{GeV}$  \cite{PDG}. Other $B_c$ mesons, such as the scalar, vector, axialvector, tensor $B_c$ mesons, have not been observed yet, but they are expected to  be produced at the  Large Hadron Collider (LHC) in the future  \cite{BC-produce,LHC}.

The  heavy quarkonium states and triply-heavy baryon states  play an important role both in studying the interplays between the perturbative and nonperturbative QCD
  and in understanding the heavy quark dynamics due to the absence  of the light quark contaminations.
The bottom-charm  quarkonium states $B_c$, which consist of the heavy quarks
with different flavors, are of special interesting. The ground states and the excited states lying below the $BD$, $BD^*$, $B^*D$, $B^*D^*$ thresholds cannot annihilate into gluons, and therefore  are more stable than the corresponding charmonium and bottomonium states, and would have widths less than a hundred $\rm{KeV}$ \cite{GI}.
The excited states  can  undergo radiative or hadronic transitions
to the ground state pseudoscalar $B_c$ mesons, which decay weakly. There have been several theoretical works on the mass spectroscopy of the $B_c$ mesons, such as
   the relativized (or relativistic) quark model with an special phenomenological potential \cite{GI,EFG,GJ,ZVR}, the  nonrelativistic quark model with  an special phenomenological potential \cite{Fulc,GKLT,EQ,Kiselev04},  the semi-relativistic quark model using the shifted large-$N$ expansion \cite{IS2004}, the perturbative QCD \cite{BV2000}, the nonrelativistic renormalization group \cite{Penin2004}, the lattice QCD \cite{Latt,Latt0909}, etc.

The QCD sum rules is a powerful theoretical tool  in
 studying   the heavy quarkonium states \cite{SVZ79,Reinders85}, and the existing works focus on the $S$-wave heavy quarkonium states $J/\psi$, $\eta_c$, $\Upsilon$, $\eta_b$, and the $P$-wave spin-triplet heavy quarkonium states $\chi_{cj}$, $\chi_{bj}$, $j=0,1,2$  \cite{Reinders85,NarisonBook}. The pseudoscalar $B_c$ mesons have been studied by  the full QCD sum rules \cite{Bagan1994,Chabab1994,Colangelo1993,Narison1988} and the potential approach
combined with the QCD sum rules \cite{GKLT,Kiselev2000,Kiselev1993}, while the vector $B_c$ mesons ($B_c^*$) have been studied by the full QCD sum rules \cite{Colangelo1993,Narison1988}, and the axialvector $B_c$ mesons have not been studied yet.
In Ref.\cite{Colangelo1993}, Colangelo,  Nardulli and Paver took the leading-order approximation,  obtained the values $M_{B_c}\approx 6.35\,\rm{GeV}$ and $f_{B_c^*}\approx f_{B_c}=(360\pm60)\,\rm{MeV}$, and did not present the value $M_{B_c^*}$. In Ref.\cite{Narison1988}, Narison took into account the
next-to-leading-order perturbative contributions  by assuming that one quark had zero mass,  and obtained the values $M_{B_c^*}-M_{B^*}=(1.53\pm0.18)\,\rm{GeV}$, $f_{B_c^*}=\frac{\sqrt{2}M_{B_c^*}}{2\gamma_{B_c^*}}$, $\gamma_{B_c^*}=14.0\pm 1.0$, the predicted mass
$M_{B_c^*}=M_{B^*}+(1.53\pm0.18)\,\rm{GeV}$ is much larger than  other theoretical calculations \cite{GI,EFG,GJ,ZVR,Fulc,GKLT,EQ,Penin2004,Latt}.
Those studies based on the QCD sum rules were preformed before the pseudoscalar $B_c$ mesons were observed by the CDF collaboration, the predictions should be updated. Now we can take the experimental data as guides to choose the suitable Borel parameters and continuum threshold parameters. Naively, we expect that the masses of the pseudoscalar, vector and axialvector $B_c$ mesons have the hierarchy: $M_{B_c(0^-)}<M_{B_c(1^-)}<M_{B_c(1^+)}$, the $0^-$, $1^-$ and $1^+$ denote the  spin-parity $J^P$. Furthermore, the calculations based on the nonrelativistic renormalization group indicate that $M_{B_c(1^-)}-M_{B_c(0^-)}=(50 \pm 17  {}^{+15}_{-12})\,\rm{MeV}$
 \cite{Penin2004}. In this article, we carry out the operator product expansion by including the next-to-leading-order perturbative contributions,  study the masses and decay constants of the vector and axialvector $B_c$ mesons with the QCD sum rules, and make reasonable predictions for the masses and decay constants, furthermore, we calculate the leptonic decay widths. The decay constants are basic input parameters in studying the exclusive  processes  of the $B_c$ mesons with the three-point vacuum correlation functions.

The article is arranged as follows:  we derive the QCD sum rules for
the masses and decay constants of the vector and axialvector $B_c$ mesons  in Sect.2;
in Sect.3, we present the numerical results and discussions; and Sect.4 is reserved for our
conclusions.

\section{QCD sum rules for  the vector and axialvector $B_c$ mesons }
In the following, we write down  the two-point correlation functions
$\Pi_{\mu\nu}(p)$  in the QCD sum rules,
\begin{eqnarray}
\Pi_{\mu\nu}(p)&=&i\int d^4x e^{ip \cdot x} \langle
0|T\left\{J_{\mu}(x)J_{\nu}^{\dagger}(0)\right\}|0\rangle \, , \\
J^V_{\mu}(x)&=&\bar{c}(x)\gamma_{\mu} b(x)  \, ,\nonumber\\
J^A_{\mu}(x)&=&\bar{c}(x)\gamma_{\mu}\gamma_5 b(x)  \, ,
\end{eqnarray}
where $J_{\mu}(x)=J^V_{\mu}(x),J^A_{\mu}(x)$, the vector and axialvector currents $J^V_\mu(x)$ and $J^A_{\mu}(x)$ interpolate the vector and axialvector $B_c$ mesons, respectively.

We can insert  a complete set of intermediate hadronic states with
the same quantum numbers as the current operators $J_{\mu}(x)$ into the
correlation functions $\Pi_{\mu\nu}(p)$  to obtain the hadronic representation
\cite{SVZ79,Reinders85}. After isolating the ground state
contributions come from the vector and axialvector $B_c$ mesons, we get the following result,
\begin{eqnarray}
\Pi_{\mu\nu}(p)&=&\frac{f_{B_c}^2M_{B_c}^2}{M_{B_c}^2-p^2}\left(-g_{\mu\nu} +\frac{p_\mu p_\nu}{p^2}\right) +\cdots\nonumber\\
&=&\Pi(p)\left(-g_{\mu\nu} +\frac{p_\mu p_\nu}{p^2}\right) +\cdots \,  ,
\end{eqnarray}
where the  decay constants $f_{B_c}$ are defined by
\begin{eqnarray}
\langle 0|J_{\mu}(0)|B_c(p)\rangle&=&f_{B_c}M_{B_c}\varepsilon_\mu \, ,
\end{eqnarray}
and the $\varepsilon_\mu$ are the  polarization vectors of the vector and axialvector $B_c$  mesons.
We can use dispersion relation to express the hadronic (or phenomenological) representation of the correlation functions $\Pi(p)$ in the following form,
\begin{eqnarray}
\Pi(p)&=&\int_{(m_b+m_c)^2}^{s_0} ds \frac{1}{s-p^2}f_{B_c}^2M_{B_c}^2 \delta(s-M_{B_c}^2)+\cdots \, ,
\end{eqnarray}
where the $s_0$ are the continuum threshold parameters.

Now, we briefly outline  the operator product
expansion for the correlation functions $\Pi_{\mu\nu}(p)$.  We contract the quark fields in the correlation functions
$\Pi^{V/A}_{\mu\nu}(p)$ (here we add the indexes $V$ and $A$ to denote the vector and axialvector currents respectively) with Wick theorem firstly,
\begin{eqnarray}
\Pi^V_{\mu\nu}(p)&=&-i\int d^4x e^{ip \cdot x}   {\rm Tr} \left\{\gamma_{\mu}B_{ij}(x)\gamma_{\nu} C_{ji}(-x) \right\}\, ,\nonumber\\
 \Pi^A_{\mu\nu}(p)&=&-i\int d^4x e^{ip \cdot x}  {\rm Tr} \left\{\gamma_{\mu}\gamma_5 B_{ij}(x)\gamma_{\nu}\gamma_5 C_{ji}(-x) \right\}\, ,\nonumber
\end{eqnarray}
where the $B_{ij}(x)$ and $C_{ij}(x)$ are the full $b$ and $c$ quark propagators, and can be written as $S_{ij}(x)$ collectively,
\begin{eqnarray}
S_{ij}(x)&=&\frac{i}{(2\pi)^4}\int d^4k e^{-ik \cdot x} \left\{
\frac{\delta_{ij}}{\!\not\!{k}-m_Q}
-\frac{g_sG^n_{\alpha\beta}t^n_{ij}}{4}\frac{\sigma^{\alpha\beta}(\!\not\!{k}+m_Q)+(\!\not\!{k}+m_Q)
\sigma^{\alpha\beta}}{(k^2-m_Q^2)^2}+\frac{\delta_{ij}\langle g^2_sGG\rangle }{12}\right.\nonumber\\
&&\left. \frac{m_Qk^2+m_Q^2\!\not\!{k}}{(k^2-m_Q^2)^4}
+\frac{g_s D_\alpha G^n_{\beta\lambda}t^n_{ij}}{3}\frac{(\!\not\!{k}+m_Q)(f^{\lambda\beta\alpha}+f^{\lambda\alpha\beta}) (\!\not\!{k}+m_Q)}{(k^2-m_Q^2)^4}+\cdots\right\} \, ,\nonumber\\
f^{\lambda\alpha\beta}&=&\gamma^\lambda(\!\not\!{k}+m_Q)\gamma^\alpha(\!\not\!{k}+m_Q)\gamma^\beta\, ,
\end{eqnarray}
and  $t^n=\frac{\lambda^n}{2}$, the $\lambda^n$ is the Gell-Mann matrix, the $i$, $j$ are color indexes, $D_\alpha=\partial_\alpha-ig_sG^n_\alpha t^n$,
and the $\langle g^2_sGG\rangle=\langle g^2_sG_{n}^{\alpha\beta}G^n_{\alpha\beta}\rangle$
is the gluon condensate \cite{Reinders85}; then complete  the integrals both in
the coordinate space and in the
momentum space, which corresponds  to calculate the  Feynman diagrams in Figs.1-3;  finally obtain the correlation functions $\Pi_{\mu\nu}(p)$ (or $\Pi(p)$) at the
level of the quark-gluon degrees  of freedom.

In calculations, we have used the equation of motion, $D^{\nu}G_{\mu\nu}^a=\sum_{q=u,d,s}g_s\bar{q}\gamma_{\mu}t^a q $,   and taken  the approximation  $\langle\bar{s}s\rangle=\langle\bar{q}q\rangle$ to obtain the contributions of the four-quark condensates.
The contributions of the four-quark condensates are depressed by inverse  powers of the large Euclidean momentum $-p^2$ (thereafter the Borel parameter $T^2$) and play  minor important roles,  we neglect other diagrams contribute to the four-quark condensates of the order $\mathcal{O}(\alpha_s^2)$. We also neglect the contributions come from the three gluon condensates, as they are also depressed by inverse powers of the large Euclidean momentum $-p^2$.

The Feynman diagrams for the next-to-leading-order perturbative contributions are shown in Fig.4.
We   calculate the diagrams using the Cutkosky's rule to obtain the   spectral densities.
There are two routines in application of  the Cutkosky's rule (or optical theorem), we resort to the routine used in Ref.\cite{Reinders85}, not the one used in Ref.\cite{Cut}.
There are ten possible cuts, the six cuts shown in Fig.5  attribute to virtual gluon emissions and correspond to the self-energy corrections and vertex corrections, while the four cuts shown in Fig.6 correspond to real gluon emissions, for technical details, one can consult Ref.\cite{Wang2013}.

\begin{figure}
 \centering
 \includegraphics[totalheight=3cm,width=5cm]{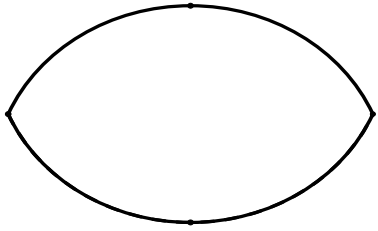}
    \caption{The leading-order perturbative contribution to the correlation functions. }
\end{figure}

\begin{figure}
 \centering
 \includegraphics[totalheight=2.7cm,width=14cm]{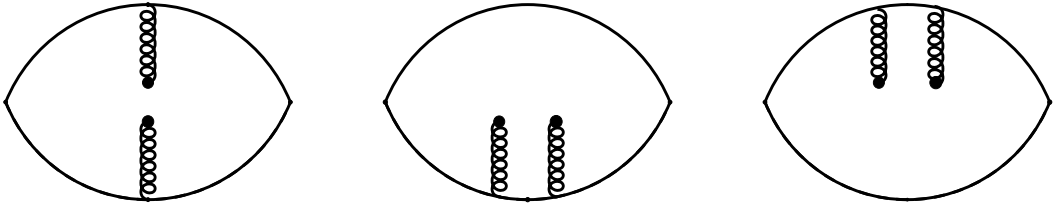}
    \caption{The diagrams contribute to the gluon condensates. }
\end{figure}
\begin{figure}
 \centering
 \includegraphics[totalheight=3cm,width=5cm]{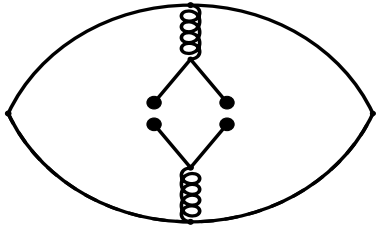}
    \caption{The typical diagram contributes to the four-quark condensate $\langle\bar{q} q\rangle^2$. }
\end{figure}

\begin{figure}
 \centering
 \includegraphics[totalheight=2.7cm,width=14cm]{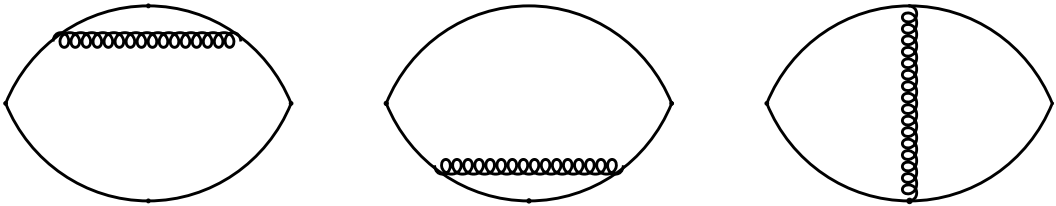}
    \caption{The next-to-leading order perturbative contributions to the correlation functions. }
\end{figure}

\begin{figure}
\centering
 \includegraphics[totalheight=3.7cm,width=14cm]{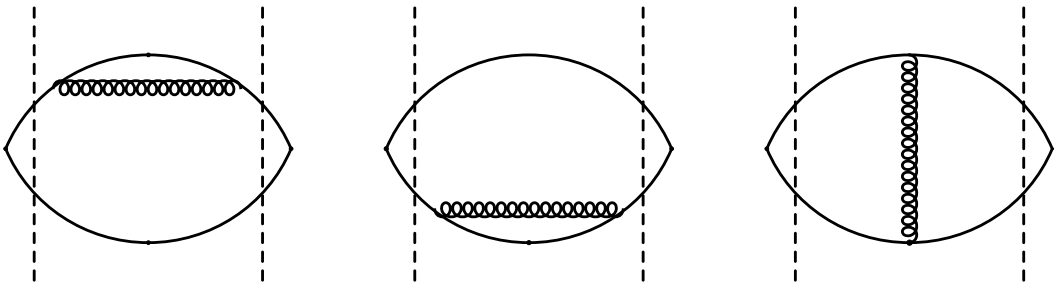}
    \caption{Six possible cuts correspond to virtual gluon emissions. }
\end{figure}

\begin{figure}
 \centering
 \includegraphics[totalheight=3.7cm,width=14cm]{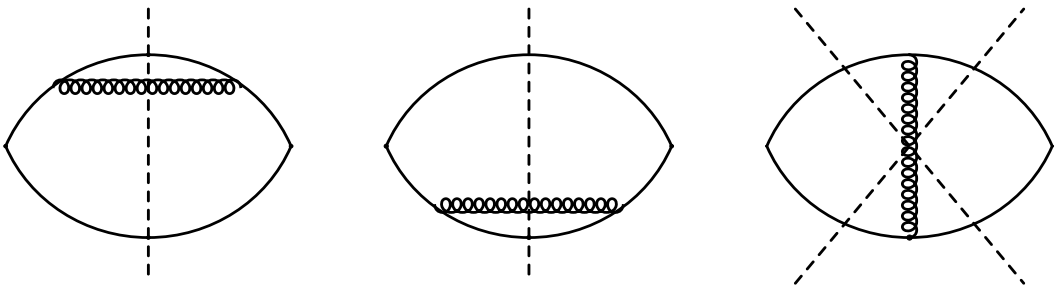}
    \caption{Four possible cuts correspond to  real gluon emissions. }
\end{figure}

 Once analytical expressions of the spectral densities at the quark level are obtained, then we take the
quark-hadron duality and perform the Borel transforms  with respect to the variable
$P^2=-p^2$ to obtain the following QCD sum rules,
\begin{eqnarray}
f_{B_c(1^{\mp})}^2 M_{B_c(1^{\mp})}^2 \exp\left(-\frac{M_{B_c(1^{\mp})}^2}{T^2}\right)&=& \int_{(m_b+m_c)^2}^{s_0} ds\left[\rho^0_{\pm}(s)+\rho^{1}_{\pm}(s)+\rho^{\rm con}_{\pm}(s)\right] \exp\left(-\frac{s}{T^2}\right) \, ,
\end{eqnarray}
where
\begin{eqnarray}
 \rho^0_{\pm}(s)&=&\frac{3}{8\pi^2}\frac{\sqrt{\lambda(s,m_b^2,m_c^2)}}{s}\left\{s-(m_b\mp m_c)^2-\frac{\lambda(s,m_b^2,m_c^2)}{3s} \right\}\, ,
\end{eqnarray}

\begin{eqnarray}
\rho_{\pm}^1(s)&=&\frac{4}{3}\frac{\alpha_s}{\pi}\rho_{\pm}^0(s)\left\{ \frac{1}{2}\overline{f}_{\pm}(s)-\overline{R}_{11}(s)-\overline{R}_{22}(s)+(s-m_b^2-m_c^2)\overline{R}_{12}(s)-\frac{5}{6}\right. \nonumber\\
&&+2\log\frac{\sqrt[4]{m_b^7m_c^7s}}{\lambda(s,m_b^2,m_c^2)}+\frac{2(s-m_b^2-m_c^2)}{\sqrt{\lambda(s,m_b^2,m_c^2)}}\log\left(\frac{1+\omega}{1-\omega}\right)
\log\frac{\lambda(s,m_b^2,m_c^2)}{m_b m_c s}\nonumber\\
&&\left. -\frac{2(s-m_b^2-m_c^2)}{3\sqrt{\lambda(s,m_b^2,m_c^2)}}\log\left(\frac{1+\omega}{1-\omega}\right)-R_{12}^1(s)\right\} \nonumber\\
&&+\frac{4}{3}\frac{\alpha_s}{\pi}\left\{ \frac{s-(m_b\mp m_c)^2}{4s\pi^2}\left[\log\left(\frac{1+\omega}{1-\omega}\right)(s-m_b^2-m_c^2)-\sqrt{\lambda(s,m_b^2,m_c^2)}\right] \right. \nonumber\\
&&+\frac{\sqrt{\lambda(s,m_b^2,m_c^2)}}{16s\pi^2}R_{12}^2(s)\left[2+\frac{(m_b\mp m_c)^2}{s}\right]-\frac{1}{\pi^2}R_0(s)+\frac{\sqrt{\lambda(s,m_b^2,m_c^2)}^3}{s^2}\nonumber\\
&&\left.\left[\frac{1}{12\pi^2} \left(1-\frac{s-m_b^2-m_c^2}{\sqrt{\lambda(s,m_b^2,m_c^2)}}\log\left(\frac{1+\omega}{1-\omega}\right) \right) -\frac{(m_b\pm m_c)\left(f_{1\pm}(s)+f_{2\pm}(s) \right)}{32\pi^2} \right]\right\} \, ,
\end{eqnarray}

\begin{eqnarray}
\rho^{\rm con}_{\pm}(s)&=&\mp \frac{m_b m_c}{24T^4}\langle\frac{\alpha_sGG}{\pi}\rangle\int_0^1dx
\left[ \frac{m_c^2}{x^3}+\frac{m_b^2}{(1-x)^3}\right]\delta(s-\widetilde{m}_Q^2) \nonumber\\
&&\pm \frac{m_b m_c}{8T^2}\langle\frac{\alpha_sGG}{\pi}\rangle\int_0^1dx\left[ \frac{1}{x^2}+\frac{1}{(1-x)^2}\right]  \delta(s-\widetilde{m}_Q^2)\nonumber\\
&&-\frac{s}{24T^4}\langle\frac{\alpha_sGG}{\pi}\rangle\int_0^1dx
\left[ \frac{(1-x)m_c^2}{x^2}+\frac{x m_b^2}{(1-x)^2}\right]\delta(s-\widetilde{m}_Q^2) \nonumber\\
&&-\frac{1}{12}\langle\frac{\alpha_sGG}{\pi}\rangle\int_0^1dx
\left[ 1+\frac{s}{2T^2}\right]\delta(s-\widetilde{m}_Q^2)+\frac{4\alpha_s^2\langle\bar{q}q\rangle^2}{81T^2}\int_0^1dx \left[\frac{2}{x(1-x)}+\frac{m_b^2 m_c^2}{x^2(1-x)^2T^4}  \right.\nonumber\\
&&\left.+\frac{m_b^2+ m_c^2\pm 9m_b m_c}{3x(1-x)T^2} -\frac{2}{3}\left(1+\frac{s}{T^2}-\frac{s^2}{T^4} \right)+\frac{5s}{3x(1-x)T^2} \right]\delta(s-\widetilde{m}_Q^2) \, ,
\end{eqnarray}
where
\begin{eqnarray}
\overline{f}_{\pm}(s)&=& \overline{V}(s)+2(s-m_b^2-m_c^2)\left[\overline{V}_{00}(s)-V_{10}(s)-V_{01}(s)+V_{11}(s)\right] \pm 2m_bm_c\nonumber\\
&&\left[ V_{10}(s)+V_{01}(s)\right]+2m_b^2\left[ V_{10}(s)-V_{20}(s)\right]+2m_c^2\left[ V_{01}(s)-V_{02}(s)\right]\, ,\nonumber\\
f_{1\pm}(s)&=&4m_bV_{20}(s)\mp 4m_cV_{01}(s)\pm 4m_cV_{11}(s)\, ,\nonumber\\
f_{2\pm}(s)&=&\pm 4m_cV_{02}(s)-4m_bV_{10}(s)+4m_bV_{11}(s)\, , \nonumber\\
\omega&=&\sqrt{\frac{s-(m_b+m_c)^2}{s-(m_b-m_c)^2}}\, ,
\end{eqnarray}
$\lambda(s,m_b^2,m_c^2)=s^2+m_b^4+m_c^4-2sm_b^2-2sm_c^2-2m_b^2m_c^2$, $\widetilde{m}_Q^2=\frac{m_b^2}{1-x}+\frac{m_c^2}{x}$, and the $T^2$ is the Borel parameter. The explicit expressions of the $\overline{V}(s)$, $\overline{V}_{00}(s)$, $V_{10}(s)$, $V_{01}(s)$, $V_{11}(s)$, $V_{20}(s)$, $V_{02}(s)$, $R_0(s)$, $\overline{R}_{11}(s)$, $\overline{R}_{22}(s)$, $\overline{R}_{12}(s)$, $R^1_{12}(s)$ and $R^2_{12}(s)$ are given in the appendix.

 We can eliminate the decay constants $f_{B_c(1^{\mp})}$ and obtain the QCD sum rules for the masses of the  vector and axialvector $B_c$ mesons,
 \begin{eqnarray}
 M_{B_c(1^{\mp})}^2&=& \frac{\int_{(m_b+m_c)^2}^{s_0} ds\frac{d}{d \left(-1/T^2\right)}\left[\rho^0_{\pm}(s)+\rho^{1}_{\pm}(s)+\rho^{\rm con}_{\pm}(s)\right]\exp\left(-\frac{s}{T^2}\right)}{\int_{(m_b+m_c)^2}^{s_0} ds \left[\rho^0_{\pm}(s)+\rho^{1}_{\pm}(s)+\rho^{\rm con}_{\pm}(s)\right]\exp\left(-\frac{s}{T^2}\right)}\,   ,
\end{eqnarray}
then use the resulting masses as input parameters to obtain the decay constants $f_{B_c(1^{\mp})}$.

\section{Numerical results and discussions}
The  mass of the  pseudoscalar $B_c$ meson is  $M_{B_c}=(6.277 \pm 0.006)\,\rm{GeV}$  from the Particle Data Group \cite{PDG}, while the calculations based on  the nonrelativistic renormalization group indicate that $M_{B_c(1^-)}-M_{B_c(0^-)}=(50 \pm 17  {}^{+15}_{-12})\,\rm{MeV}$
 \cite{Penin2004}. We can tentatively take
  the continuum threshold parameters as $s^0_{B_c(1^-)}=(41-47)\,\rm{GeV}^2$ and $s^0_{B_c(1^+)}=(46-54)\,\rm{GeV}^2$, and search for the ideal values, where we have assumed that an additional $P$-wave results in  mass-shift $0.5\,\rm{GeV}$ and the energy gap between the ground states and the first radial excited states is $0.5\,\rm{GeV}$.

The quark condensate is taken to be the standard value
$\langle\bar{q}q \rangle=-(0.24\pm0.01\, \rm{GeV})^3$ at the energy scale  $\mu=1\, \rm{GeV}$ \cite{ColangeloReview}. The quark condensate evolves with the   renormalization group equation, $\langle\bar{q}q \rangle(\mu^2)=\langle\bar{q}q \rangle(Q^2)\left[\frac{\alpha_{s}(Q)}{\alpha_{s}(\mu)}\right]^{\frac{4}{9}}$.
 The value of the gluon condensate $\langle \frac{\alpha_s
GG}{\pi}\rangle $ has been updated from time to time, and changes
greatly \cite{NarisonBook}, we use the recently updated value $\langle \frac{\alpha_s GG}{\pi}\rangle=(0.022 \pm
0.004)\,\rm{GeV}^4 $ \cite{gg-conden,Narison-gc-1105}.

In this article, we study the vector and axialvector $B_c$ mesons with both the $\overline{MS}$ masses and pole masses.
The $\overline{MS}$  masses  have been studied extensively  by the QCD sum rules and Lattice QCD \cite{PDG,NarisonBook,ColangeloReview}. The values listed  in the Review of Particle Physics are $\overline{m}_c(\overline{m}_c^2)=1.275\pm 0.025\,\rm{GeV}$ and $\overline{m}_b(\overline{m}_b^2)=4.18 \pm0.03\,\rm{GeV}$ \cite{PDG},
 which correspond to the pole masses $m_c=(1.67 \pm 0.07)\, \rm GeV$ and  $m_b=(4.78 \pm 0.06)\, \rm GeV$.  The recent studies based on the QCD sum rules  \cite{Narison-gc-1105,Chetyrkin-2009}, the
 nonrelativistic large-n $\Upsilon$ sum rules with renormalization group improvement \cite{Hoang-2012}  and the lattice QCD \cite{McNeile-latt} indicate (slightly) different values. We take the $\overline{MS}$ masses $\overline{m}_{c}(\overline{m}_c^2)=(1.275\pm0.025)\,\rm{GeV}$ and  $\overline{m}_{b}(\overline{m}_b^2)=(4.18\pm 0.03)\,\rm{GeV}$
 from the Particle Data Group \cite{PDG}. Furthermore, we take into account
the energy-scale dependence of  the $\overline{MS}$ masses from the renormalization group equation,
\begin{eqnarray}
\overline{m}_c(\mu^2)&=&\overline{m}_c(\overline{m}_c^2)\left[\frac{\alpha_{s}(\mu)}{\alpha_{s}(\overline{m}_c)}\right]^{\frac{12}{25}} \, ,\nonumber\\
\overline{m}_b(\mu^2)&=&\overline{m}_b(\overline{m}_b^2)\left[\frac{\alpha_{s}(\mu)}{\alpha_{s}(\overline{m}_b)}\right]^{\frac{12}{23}} \, ,\nonumber\\
\alpha_s(\mu)&=&\frac{1}{b_0t}\left[1-\frac{b_1}{b_0^2}\frac{\log t}{t} +\frac{b_1^2(\log^2{t}-\log{t}-1)+b_0b_2}{b_0^4t^2}\right]\, ,
\end{eqnarray}
  where $t=\log \frac{\mu^2}{\Lambda^2}$, $b_0=\frac{33-2n_f}{12\pi}$, $b_1=\frac{153-19n_f}{24\pi^2}$, $b_2=\frac{2857-\frac{5033}{9}n_f+\frac{325}{27}n_f^2}{128\pi^3}$,  $\Lambda=213\,\rm{MeV}$, $296\,\rm{MeV}$  and  $339\,\rm{MeV}$ for the flavors  $n_f=5$, $4$ and $3$, respectively  \cite{PDG}.

\begin{figure}
 \centering
 \includegraphics[totalheight=6cm,width=7cm]{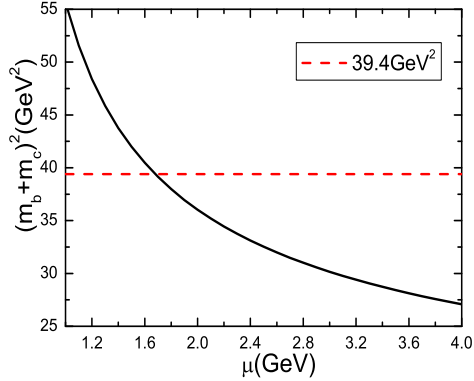}
        \caption{ The energy scale dependence of the threshold $(\overline{m}_b(\mu^2)+\overline{m}_c(\mu^2))^2$, where $39.4\,\rm{GeV}^2$ is the squared mass of the pseudoscalar meson $B_c$.}
\end{figure}

 In Fig.7, we plot the threshold $(\overline{m}_b+\overline{m}_c)^2$ with variations of the energy scales. From the figure, we can see that the threshold $(\overline{m}_b+\overline{m}_c)^2$ decreases quickly with increase of the energy scale,  the energy scale should be larger than $1.7\,\rm{GeV}$ for the $b\bar{c}$ or $c\bar{b}$ system, we can take the typical energy scale  $\mu=2\,\rm{GeV}$, which corresponds to the threshold $(\overline{m}_b+\overline{m}_c)^2\approx 36.0\,\rm{GeV}^2$.
On the other hand, if we take the pole masses $m_c=1.67 \, \rm GeV$ and  $m_b=4.78 \, \rm GeV$ from the Particle Data Group \cite{PDG}, the threshold $(m_b+m_c)^2=41.6\,\rm{GeV}^2$ is larger than the value $39.4\,\rm{GeV}^2$ of  the squared mass of the pseudoscalar meson $B_c$. We have to choose much smaller values, $m_c=1.3\,\rm{GeV}$ and $m_b=4.7\,\rm{GeV}$, which corresponds to the threshold $(m_b+m_c)^2=36.0\,\rm{GeV}^2$. Furthermore, we  choose the uncertainties   as that of the  $\overline{MS}$ masses from the Particle Data Group tentatively \cite{PDG}.

The pole masses and the $\overline{MS}$ masses
have the relation  $m_Q
=\overline{m}_Q(\overline{m}_Q^2)\left[1+\frac{4 \alpha_s(\overline{m}_Q^2)}{3\pi}+\cdots\right]$, we maybe expect that a simple replacement of the corresponding quantities in the spectral densities $\rho^0(s)$, $\rho^1(s)$ and $\rho^{\rm con}(s)$ can lead to analogous results, such an  expectation  is sensible  only in the case that the integral ranges $\int_{(m_c+m_b)^2}^{ s_0}$ and $\int_{(\overline{m}_c+\overline{m}_b)^2}^{s_0}$ are large enough, the variations $m_Q-\overline{m}_Q$ are small enough so as to  be neglected.
In the present case, the integral ranges are small, we have to fit the parameters independently.
 For the $\overline{MS}$ masses, we observe that the ideal parameters are $T^2=(5.0-7.0)\,\rm{GeV}^2$ [$(7.0-9.0)\,\rm{GeV}^2$] and $s_0=(45\pm 1)\,\rm{GeV}^2$ [$(54\pm 1)\,\rm{GeV}^2$] for the vector [axialvector] $B_c$ mesons, the corresponding pole contributions and the resulting  masses and decay constants are presented in Table 1 and Figs.8-9.
For the pole masses, we observe that the ideal parameters are $T^2=(5.4-7.4)\,\rm{GeV}^2$ [$(7.4-9.4)\,\rm{GeV}^2$] and $s_0=(45\pm 1)\,\rm{GeV}^2$ [$(54\pm 1)\,\rm{GeV}^2$] for the vector [axialvector] $B_c$ mesons, the corresponding pole contributions and the resulting  masses and decay constants are also presented in Table 1 and Figs.8-9.
The threshold parameters and predicted masses satisfy the relations $\sqrt{s^0_{B_c(1^-)}}-M_{B_c(1^-)}\approx 0.4\,\rm{GeV}$ and $\sqrt{s^0_{B_c(1^+)}}-M_{B_c(1^+)}\approx 0.6\,\rm{GeV}$, which are compatible with our naive expectation that the energy gap between the ground state and first radial excited is about $0.5\,\rm{GeV}$.
 The calculations based on the nonrelativistic renormalization group indicate that $M_{B_c(1^-)}-M_{B_c(0^-)}=(50 \pm 17  {}^{+15}_{-12})\,\rm{MeV}$ \cite{Penin2004}, the present prediction $M_{B_c(1^-)}-M_{B_c(0^-)}\approx 60\,\rm{MeV}$  is satisfactory.

\begin{table}
\begin{center}
\begin{tabular}{|c|c|c|c|c|c|c|c|}\hline\hline
                           & $T^2 (\rm{GeV}^2)$  & $s_0 (\rm{GeV}^2)$   & pole         & $M_{B_c}(\rm{GeV})$  & $f_{B_c}(\rm{GeV})$   \\ \hline
 $B_{c}({1}^-)$            & $5.0-7.0$           & $45\pm1$             & $(50-75)\%$  & $6.337\pm0.052$      & $0.384\pm0.032$         \\ \hline
 $B_{c}({1}^+)$            & $7.0-9.0$           & $54\pm1$             & $(54-73)\%$  & $6.730\pm0.061$      & $0.373\pm0.025$          \\ \hline
 $\widehat{B}_{c}({1}^-)$  & $5.4-7.4$           & $45\pm1$             & $(50-74)\%$  & $6.331\pm0.047$      & $0.415\pm0.031$         \\ \hline
 $\widehat{B}_{c}({1}^+)$  & $7.4-9.4$           & $54\pm1$             & $(52-70)\%$  & $6.737\pm0.056$      & $0.374\pm0.023$          \\ \hline
 \hline
\end{tabular}
\end{center}
\caption{ The Borel parameters, continuum threshold parameters, pole contributions, masses and decay constants  of  the vector and axialvector $B_c$ mesons. The wide-hat denotes that the pole masses are used.  }
\end{table}

In Table 2,  we present the theoretical values of the masses of  the vector and axialvector $B_c$ mesons from the relativized (or relativistic) quark model with an special potential \cite{GI,EFG,GJ,ZVR}, the nonrelativistic quark model with  an special potential \cite{Fulc,GKLT,EQ}, and the lattice QCD \cite{Latt}. From the Table, we can see that the present predictions are consistent  with those values. In Table 3, we present the values of the decay constants of the vector and axialvector $B_c$ mesons from the relativistic quark model with an special potential \cite{EFG}, the nonrelativistic quark model with  an special potential \cite{Fulc,GKLT,EQ,Kiselev04}, the light-front quark model \cite{ChoiJ,Hwang}, the Bethe-Salpeter equation \cite{WangGL}, and field correlator method \cite{BBS}. The present predictions $f_{B_c(1^-)}=0.384\pm0.032\,(0.416)$,   $0.415\pm0.031\,(0.459)\, \rm{GeV}$  are compatible  with those theoretical calculations  $f_{B_c(1^-)}= (380-520)\,\rm{MeV}$ \cite{EFG,Fulc,GKLT,EQ,Kiselev04,ChoiJ,Hwang,WangGL,BBS},  while the present prediction $f_{B_c(1^+)}=0.373\pm0.025$,
$0.374\pm0.023\,\rm{GeV}$ is much larger than the value $160\,\rm{MeV}$ from the Bethe-Salpeter equation \cite{WangGL}. At present time, it is difficult to say which value is  superior to others.

\begin{figure}
 \centering
 \includegraphics[totalheight=5cm,width=6cm]{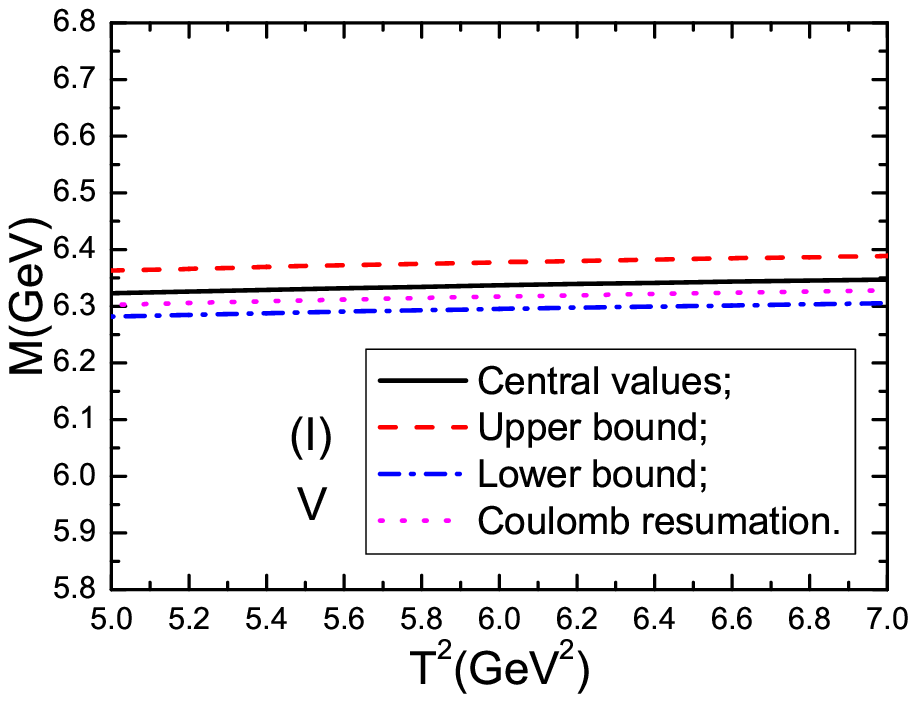}
\includegraphics[totalheight=5cm,width=6cm]{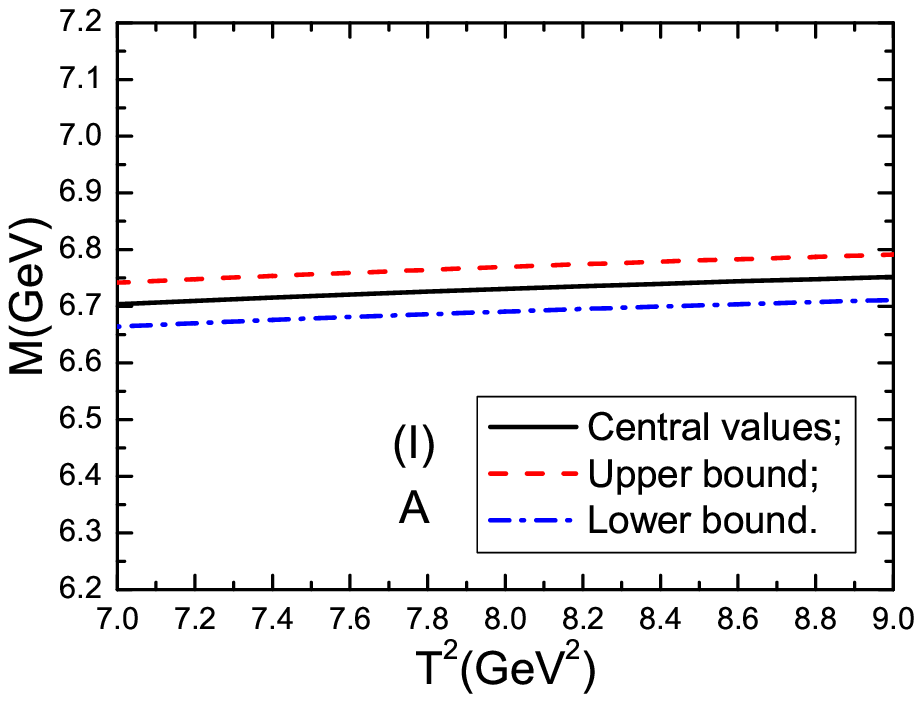}
\includegraphics[totalheight=5cm,width=6cm]{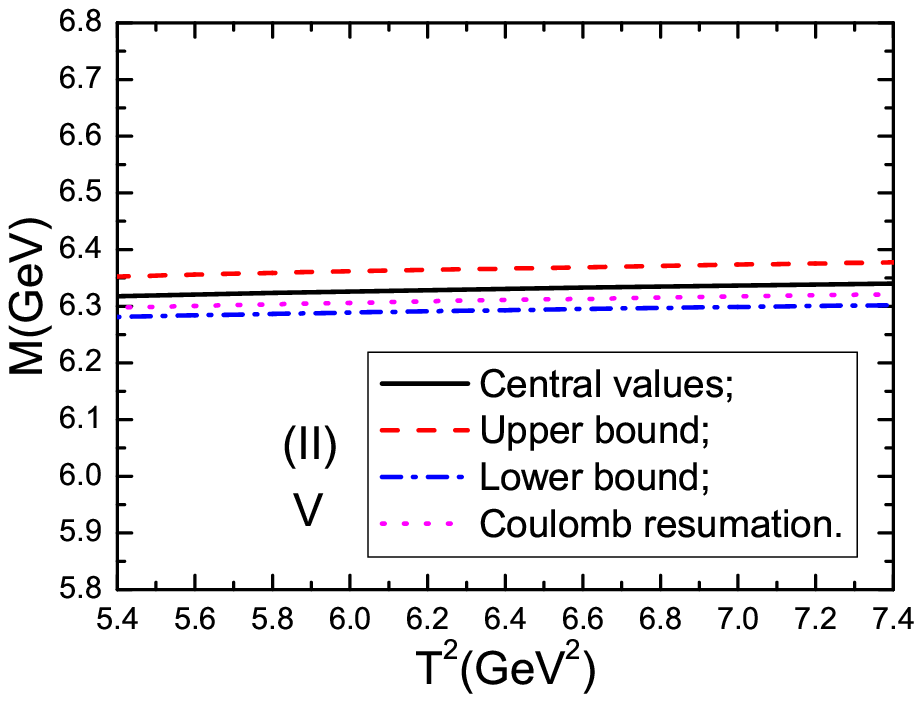}
\includegraphics[totalheight=5cm,width=6cm]{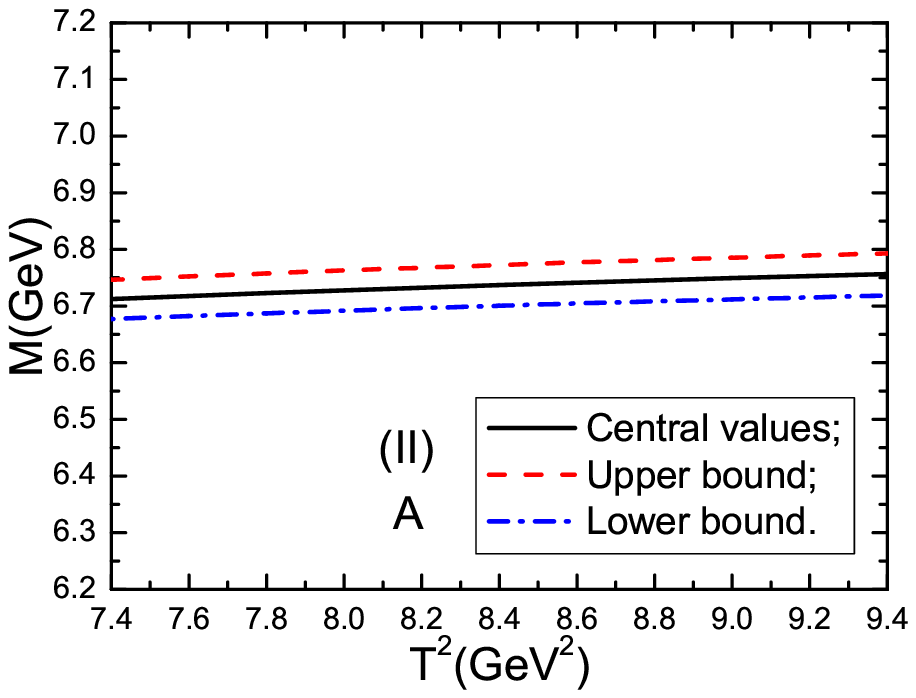}
        \caption{ The masses  of the vector ($V$) and axialvector ($A$) $B_c$ mesons with variations of the Borel parameters $T^2$. In (I) and (II),  we use  the $\overline{MS}$ masses and pole masses, respectively.}
\end{figure}

 The leptonic decay widths $\Gamma_{\ell \bar{\nu}_{\ell}}$ of the vector and axialvector $B_c$ mesons can be written as,
\begin{eqnarray}
\Gamma_{\ell \bar{\nu}_{\ell}} &=& \frac{G_F^2}{4\pi}|V_{bc}|^2f_{B_c}^2M_{B_c}^3 \left(1-\frac{M_{\ell}^2}{M_{B_c}^2} \right)^2\left(1+\frac{M_{\ell}^2}{2M_{B_c}^2} \right)\, ,
\end{eqnarray}
where $\ell=e,\mu,\tau$, the Fermi constant $G_F=1.16637\times10^{-5} \,\rm{GeV}^{-2}$, the CKM matrix element  $V_{cb}=40.6\times 10^{-3}$, the masses of the leptons  $m_e=0.511\times 10^{-3}\,\rm{GeV}$, $m_\mu=105.658\times 10^{-3}\,\rm{GeV}$,
$m_\tau=1776.82\times 10^{-3}\,\rm{GeV}$ \cite{PDG}. We use the masses and decay constants of the vector and axialvector $B_c$ mesons come from the $\overline{MS}$ masses (pole masses) to obtain the leptonic decay widths,
\begin{eqnarray}
\Gamma_{B_c(1^-)\to e \bar{\nu}_{e}}&=&0.670{}^{+0.016}_{-0.017} {}^{+0.116}_{-0.107} \, \left(0.780{}^{+0.017}_{-0.018} {}^{+0.121}_{-0.112}\right)\times 10^{-3}\,\rm{eV},  \nonumber \\
\Gamma_{B_c(1^-)\to \mu \bar{\nu}_{\mu}}&=&0.669{}^{+0.017}_{-0.016} {}^{+0.117}_{-0.107}\,\left(0.780{}^{+0.017}_{-0.018} {}^{+0.120}_{-0.113} \right) \times 10^{-3}\,\rm{eV},  \nonumber \\
\Gamma_{B_c(1^-)\to \tau \bar{\nu}_{\tau}}&=&0.591{}^{+0.016}_{-0.016} {}^{+0.102}_{-0.095}\,\left(0.688{}^{+0.017}_{-0.017} {}^{+0.107}_{-0.099}\right) \times 10^{-3}\,\rm{eV}, \nonumber \\
\Gamma_{B_c(1^+)\to e \bar{\nu}_{e}}&=&0.757{}^{+0.019}_{-0.021} {}^{+0.105}_{-0.098} \,\left(0.763{}^{+0.019}_{-0.019} {}^{+0.097}_{-0.091} \right)\times 10^{-3}\,\rm{eV},  \nonumber \\
\Gamma_{B_c(1^+)\to \mu \bar{\nu}_{\mu}}&=&0.757{}^{+0.020}_{-0.021} {}^{+0.104}_{-0.098} \, \left(0.763{}^{+0.019}_{-0.019} {}^{+0.097}_{-0.091}\right)\times 10^{-3}\,\rm{eV}, \nonumber \\
\Gamma_{B_c(1^+)\to \tau \bar{\nu}_{\tau}}&=&0.678{}^{+0.020}_{-0.020} {}^{+0.094}_{-0.088} \,\left(0.684{}^{+0.018}_{-0.018} {}^{+0.086}_{-0.082} \right)\times 10^{-3}\,\rm{eV} .
\end{eqnarray}
where the uncertainties originate from the  uncertainties of the  masses and decay constants, respectively.
The radiative decay widths of the electric dipole (or magnetic dipole) transitions  $B^{\pm}_c(1^+) \to B^{\pm}_c(1^-)\gamma$ (or $B^{\pm}_c(1^-) \to B^{\pm}_c\gamma$) are about tens of $\rm{KeV}$ (or $\rm eV$) from the potential models \cite{EFG,Fulc,GKLT,EQ}, the branching fractions of the $B_c(1^+)\to  \ell \bar{\nu}_\ell$ (or $B_c(1^-)\to  \ell \bar{\nu}_\ell$) are of the order $10^{-7}$ (or $10^{-4}$), the tiny (or small) branching fractions  maybe (or maybe not) escape experimental  detections.   The $b\bar{b}$ pairs and the $S$-wave, $P$-wave $B_c$ mesons would be copiously produced at the LHCb \cite{BC-produce,LHC},   we expect that a large number of vector and axialvector mesons events would be accumulated, and the experimental study of the  branching fractions of the  leptonic decays of  vector (maybe also the axialvector)  $B_c$ mesons  are feasible.

 \begin{figure}
 \centering
 \includegraphics[totalheight=5cm,width=6cm]{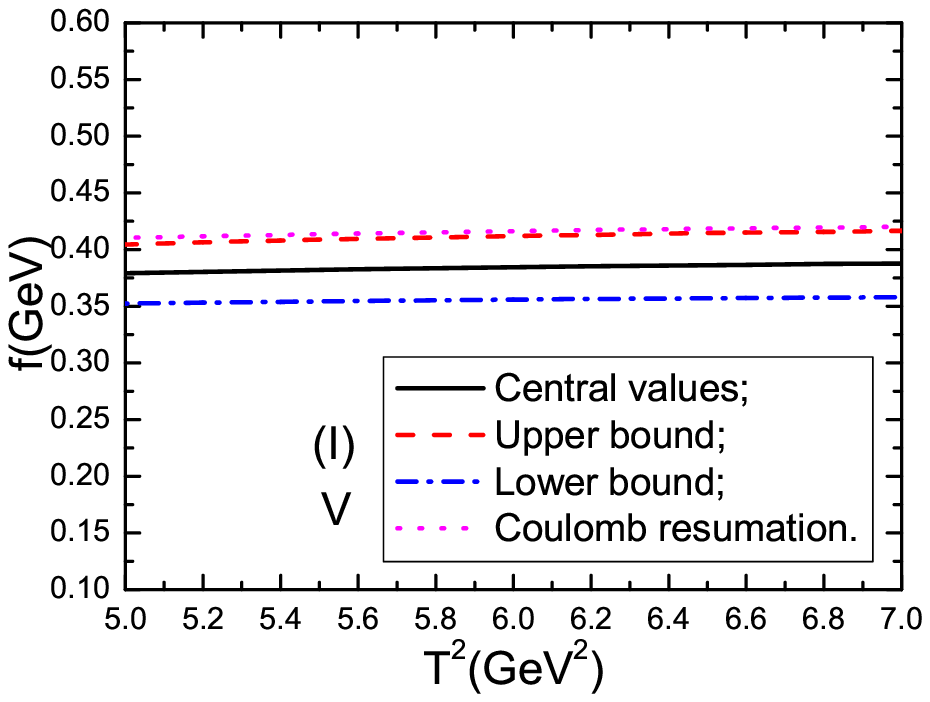}
\includegraphics[totalheight=5cm,width=6cm]{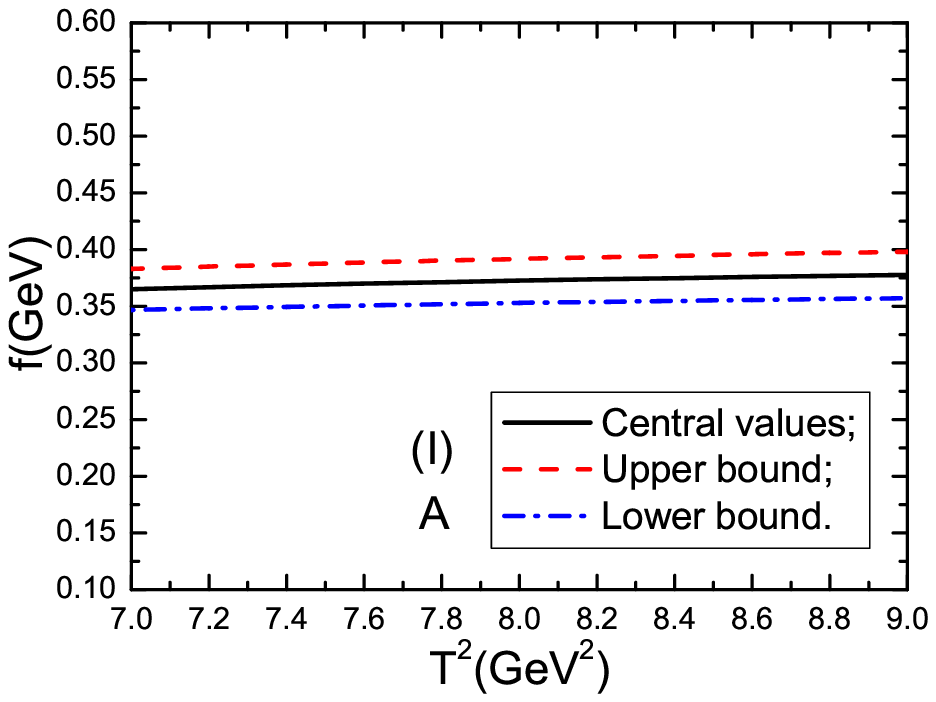}
\includegraphics[totalheight=5cm,width=6cm]{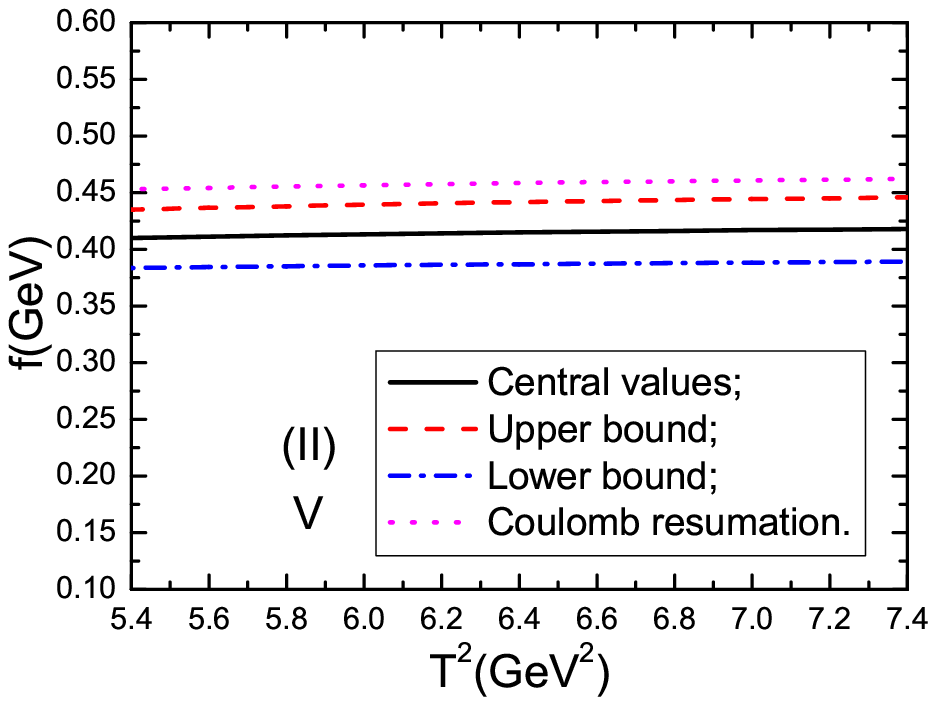}
\includegraphics[totalheight=5cm,width=6cm]{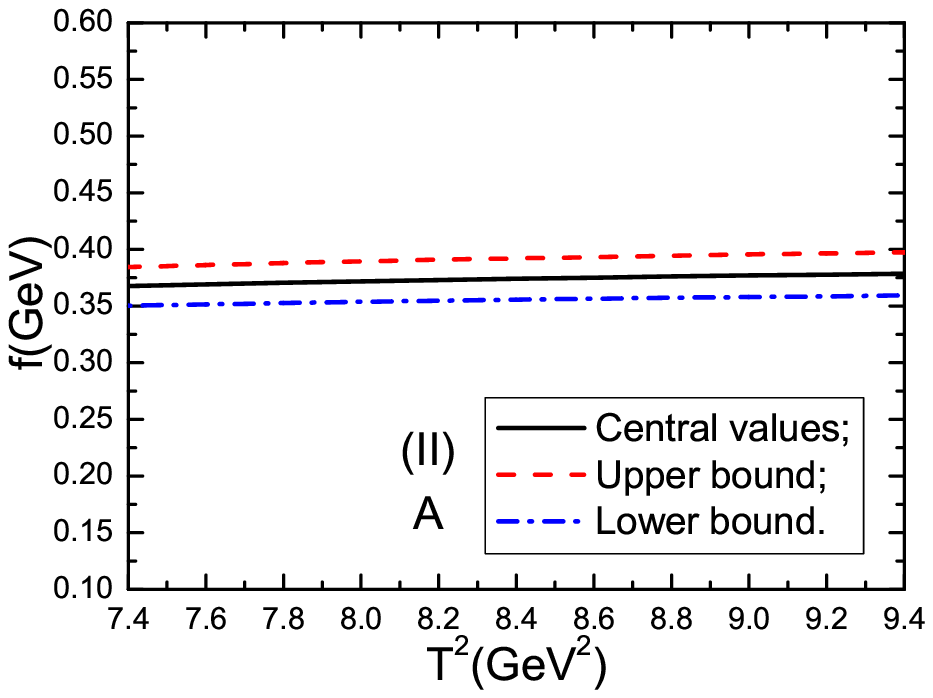}
        \caption{ The decay constants  of the vector ($V$) and axialvector ($A$) $B_c$ mesons with variations of the Borel parameters $T^2$. In (I) and (II),  we use  the $\overline{MS}$ masses and pole masses, respectively.}
\end{figure}

\begin{table}
\begin{center}
\begin{tabular}{|c|c|c|c|c|c|c|c|c|c|}\hline\hline
                  &\cite{GI}  &\cite{EFG} &\cite{GJ} &\cite{ZVR} &\cite{Fulc} &\cite{GKLT} &\cite{EQ} &\cite{Latt}  &This work      \\ \hline
 $B_{c}({1}^-)$   &6.338      &6.332      &6.308     &6.340      &6.341       &6.317       &6.337     &6.321        &$6.337\pm0.052\,(6.317)$   \\
                  &           &           &          &           &            &            &          &             &$6.331\pm0.047\,(6.311)$  \\ \hline
 $B_{c}({1}^+)$   &6.741      &6.734      &6.738     &6.730      &6.737       &6.717       &6.730     &6.743        &$6.730\pm0.061$   \\
                  &           &           &          &           &            &            &          &             &$6.737\pm0.056$   \\ \hline
 \hline
\end{tabular}
\end{center}
\caption{ The masses of the vector and axialvector $B_c$ mesons from different theoretical approaches, the unit is GeV. The values in the bracket denote the Coulomb-like corrections are taken into account.}
\end{table}

\begin{table}
\begin{center}
\begin{tabular}{|c|c|c|c|c|c|c|c|c|c|c|}\hline\hline
                &\cite{EFG} &\cite{Fulc} &\cite{GKLT} &\cite{EQ} &\cite{Kiselev04} &\cite{ChoiJ} &\cite{Hwang}&\cite{WangGL}&\cite{BBS} &This work    \\ \hline
$B_{c}({1}^-)$  &503        &517         &460         &500       &400              &398          &387         &418          &453        &$384\pm32\,(416)$  \\
                &           &            &            &          &                 &             &            &             &           &$415\pm31\,(459)$ \\\hline
 $B_{c}({1}^+)$ &           &            &            &          &                 &             &            &160          &           &$373\pm25$    \\
                &           &            &            &          &                 &             &            &             &           &$374\pm23$  \\ \hline
 \hline
\end{tabular}
\end{center}
\caption{ The decay constants of the vector and axialvector $B_c$ mesons from different theoretical approaches, the unit is MeV.
The values in the bracket denote the Coulomb-like corrections are taken into account.}
\end{table}

For the heavy quarkonium states, the relative velocities $\omega$ of the quarks  are small, we should account for the Coulomb-like $\frac{\alpha_s^\mathcal{C}}{\omega}$  corrections.  After taking into account  all the Coulomb-like contributions, we obtain the coefficient $F(\omega)$ to dress   the leading-order spectral densities  $\rho^0_{\pm}(s)$ \cite{Kiselev2000,Coulomb-BC},
\begin{eqnarray}
F(\omega)&=&\frac{4\pi\alpha_s^\mathcal{C}}{3\omega} \frac{1}{1-\exp\left(-\frac{4\pi\alpha_s^\mathcal{C}}{3\omega}\right)}=1+\frac{2\pi\alpha_s^\mathcal{C}}{3\omega}+\cdots\,   .
\end{eqnarray}
If we take the approximation $\alpha_s^{\mathcal{C}}=\alpha_s$, then
 $1+\frac{\rho^1_{+}(s)}{\rho^0_{+}(s)}\approx 1+\frac{2\pi\alpha_s}{3\omega}\ll 1+\frac{\rho^1_{-}(s)}{\rho^0_{-}(s)}$. In Fig.10, we plot the ratio $R=\frac{\rho^0_{-}(s)}{\rho^0_{+}(s)}$ of the leading-order spectral densities, where the $\overline{MS}$ masses are used.  From the figure we can see that $\rho^0_{-}(s) \ll \rho^0_{+}(s)$. The terms in the next-to-leading order spectral density $\rho^1_{-}(s)$ cannot be factorized as $\frac{4}{3}\frac{\alpha_s}{\pi}\rho^0_{-}(s)\,g(s,m_b,m_c)$ lead to the behavior $\frac{2\pi\alpha_s}{3\omega}\ll \frac{\rho^1_{-}(s)}{\rho^0_{-}(s)}$, where $g(s,m_b,m_c)$ is a formal notation.
\begin{figure}
 \centering
 \includegraphics[totalheight=5cm,width=7cm]{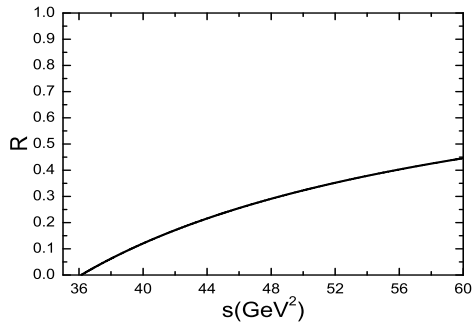}
        \caption{ The  ratio  $R=\frac{\rho^0_{-}(s)}{\rho^0_{+}(s)}$ of the leading-order spectral densities.}
\end{figure}
The  next-to-leading order spectral density $\rho^1_{+}(s)$ can be  approximated  by $\rho^0_{+}(s)\frac{2\pi\alpha_s^{\mathcal{C}}}{3\omega}$.
We  account for all the   Coulomb-like contributions by multiplying the leading-order spectral density $\rho^0_{+}(s)$ by the coefficient $F(\omega)$ tentatively, and  obtain the central values
\begin{eqnarray}
M_{B_{c}({1}^-)}&=&6.317\, (6.311) \,\rm{GeV} \, ,  \nonumber\\
f_{B_{c}({1}^-)}&=&0.416\, (0.459) \,\rm{GeV} \, ,
\end{eqnarray}
with the $\overline{MS}$ masses (pole masses), those  predictions are also shown in Tables 2-3. The mass-shifts are about $\delta M_{B_{c}({1}^-)} \approx -20\,\rm{MeV}$, while the decay constant shifts are about $\delta f_{B_{c}({1}^-)} \approx +(30-40)\,\rm{MeV}$.

\section{Conclusion}
In this article,  we study the vector and axialvector $B_c$ mesons by including the next-to-leading order perturbative contributions in the operator product expansion with the QCD sum rules,
and make reasonable predictions for the masses and decay constants, then calculate
the leptonic decay widths. The present predictions for the masses and decay constants can be confronted with the experimental data in the future at the LHC.
We can also take the masses and decay constants as basic input parameters and study  other phenomenological quantities, such as the semi-leptonic, non-leptonic and radiative  decays.

\section*{Acknowledgements}
This  work is supported by National Natural Science Foundation,
Grant Number 11075053,  and the Fundamental Research Funds for the
Central Universities.

\section*{Appendix}
The notations in the next-to-leading order spectral densities,
\begin{eqnarray}
\overline{V}_{00}(s)&=& \frac{1}{\sqrt{\lambda(s,m_b^2,m_c^2)}}\left\{ \frac{\log^2(1-\omega_1^2)}{4}-\log^2(1+\omega_1)+ \frac{\log^2(1-\omega_2^2)}{4}-\log^2(1+\omega_2) \right. \nonumber \\
&& +2\log(\omega_1+\omega_2)\log\left(\frac{1+\omega}{1-\omega}\right)-\log\omega_1\log\left(\frac{1+\omega_2}{1-\omega_2}\right)-\log\omega_2\log\left(\frac{1+\omega_1}{1-\omega_1}\right) \nonumber \\
&&\left.-{\rm Li_2}\left( \frac{2\omega_1}{1+\omega_1} \right) -{\rm Li_2}\left( \frac{2\omega_2}{1+\omega_2} \right)+\pi^2\right\} \, ,\nonumber
\end{eqnarray}

\begin{eqnarray}
V_{10}(s)&=& \frac{1}{s}\left\{\frac{1}{2}\log\left(\frac{1-\omega_1^2}{1-\omega_2^2}\right)-\frac{1}{\omega_2}\log\left(\frac{1+\omega}{1-\omega} \right)+\log\frac{\omega_2}{\omega_1}\right\} \, ,\nonumber \\
V_{01}(s)&=& V_{10}(s)|_{\omega_1 \leftrightarrow \omega_2} \, ,\nonumber
\end{eqnarray}

\begin{eqnarray}
V_{20}(s)&=& \frac{1}{2s}\left\{-\frac{\omega_1\omega_2}{\omega_1+\omega_2}\log\left(\frac{1+\omega}{1-\omega} \right)-\frac{\omega_1}{\omega_2(\omega_1+\omega_2)}\log\left(\frac{1+\omega}{1-\omega} \right)+   \frac{\omega_1}{\omega_1+\omega_2}\log\left(\frac{1-\omega_1^2}{1-\omega_2^2}\right)\right.\nonumber\\
&&\left.+\frac{2\omega_1}{\omega_1+\omega_2}\log\frac{\omega_2}{\omega_1}+1\right\} \, ,\nonumber \\
V_{02}(s)&=&V_{20}(s)|_{\omega_1 \leftrightarrow \omega_2} \, ,\nonumber
\end{eqnarray}

\begin{eqnarray}
V_{11}(s)&=& \frac{1}{2s}\left\{\frac{\omega_1\omega_2}{\omega_1+\omega_2}\log\left(\frac{1+\omega}{1-\omega} \right)-  \frac{\omega_1-\omega_2}{2(\omega_1+\omega_2)}\log\left(\frac{1-\omega_1^2}{1-\omega_2^2}\right)-\frac{1}{\omega_1+\omega_2}\log\left(\frac{1+\omega}{1-\omega} \right)\right.\nonumber\\
&&\left.+\frac{\omega_1}{\omega_1+\omega_2}\log\frac{\omega_1}{\omega_2}+\frac{\omega_2}{\omega_1+\omega_2}\log\frac{\omega_2}{\omega_1}-1\right\} \, ,\nonumber
\end{eqnarray}

\begin{eqnarray}
\overline{V}(s)&=& -\frac{2\omega_1\omega_2}{\omega_1+\omega_2}\log\left(\frac{1+\omega}{1-\omega} \right)
-\frac{\omega_2}{\omega_1+\omega_2}\log(1-\omega_1^2)-\frac{\omega_1}{\omega_1+\omega_2}\log(1-\omega_2^2)+2\log(\omega_1+\omega_2)\nonumber \\
&&-2\frac{\omega_1\log\omega_1+\omega_2\log\omega_2}{\omega_1+\omega_2}\, ,\nonumber
\end{eqnarray}

  \begin{eqnarray}
R_0(s)&=&-\frac{s m_b^2+s m_c^2-2m_b^2m_c^2}{4s}\log\left(\frac{1+\omega}{1-\omega}\right)+\frac{\sqrt{\lambda(s,m_b^2,m_c^2)}(s+m_b^2+m_c^2)}{8s}  \nonumber\\
&&+\frac{m_b^2-m_c^2}{4}\log\left(\frac{M+\omega}{M-\omega}\right)\, , \nonumber
\end{eqnarray}

\begin{eqnarray}
 \overline{R}_{11}(s)&=&-\frac{s+m_b^2-m_c^2}{2\sqrt{\lambda(s,m_b^2,m_c^2)}} \log\left(\frac{1+\omega_1}{1-\omega_1}\right) -\frac{m_b^2-m_c^2}{\sqrt{\lambda(s,m_b^2,m_c^2)}} \log\left(\frac{1+\omega_1}{1-\omega_1}\right) \nonumber\\
&&-\frac{s-m_b^2+m_c^2}{\sqrt{\lambda(s,m_b^2,m_c^2)}} \log\left(\frac{1+\omega}{1-\omega}\right) \nonumber \\
\overline{R}_{22}(s)&=&\overline{R}_{11}(s)|_{m_b\leftrightarrow m_c} \, , \nonumber
\end{eqnarray}
\begin{eqnarray}
\overline{R}_{12}(s)&=&\frac{1}{\sqrt{\lambda(s,m_b^2,m_c^2)}}
\left\{-2\log\frac{m_b}{m_c}\log\left(\frac{M+\omega}{M-\omega}\right)-\log^2\left(\frac{1+\omega}{1-\omega}\right)+2\log\frac{s}{\bar{s}}\log\left(\frac{1+\omega}{1-\omega}\right)
 \right. \nonumber \\
&&-4{\rm Li_2}\left( \frac{2\omega}{1+\omega}\right)+2{\rm Li_2}\left( \frac{\omega-1}{\omega-M}\right) +2{\rm Li_2}\left( \frac{\omega-1}{\omega+M}\right)-2{\rm Li_2}\left( \frac{\omega+1}{\omega-M}\right)-2{\rm Li_2}\left( \frac{\omega+1}{\omega+M}\right)\nonumber\\
&&-\frac{1}{2}{\rm Li_2}\left( \frac{1+\omega_1}{2}\right)-\frac{1}{2}{\rm Li_2}\left( \frac{1+\omega_2}{2}\right)-{\rm Li_2}\left( \omega_1\right)-{\rm Li_2}\left( \omega_2\right)+\frac{\log2\log\left[(1+\omega_1)(1+\omega_2) \right]}{2}\nonumber\\
&&\left.-\frac{\log^2 2}{2}+\frac{\pi^2}{12}\right\}\, ,\nonumber
\end{eqnarray}
\begin{eqnarray}
R^1_{12}(s)&=&\frac{s}{\sqrt{\lambda(s,m_b^2,m_c^2)}} \left\{\log^2(1-\omega)-\log^2(1+\omega) +2\log\frac{2s}{\bar{s}}\log\left(\frac{1+\omega}{1-\omega}\right)
+2{\rm Li_2}\left(\frac{1-\omega}{2}\right)\right. \nonumber \\
&&\left.-2{\rm Li_2}\left(\frac{1+\omega}{2}\right)+2{\rm Li_2}\left(\frac{1+\omega}{1+M}\right)+2{\rm Li_2}\left(\frac{1+\omega}{1-M}\right)
-2{\rm Li_2}\left(\frac{1-\omega}{1-M}\right)-2{\rm Li_2}\left(\frac{1-\omega}{1+M}\right)\right\} \, ,\nonumber
\end{eqnarray}

\begin{eqnarray}
R^2_{12}(s)&=&\frac{s^2}{\sqrt{\lambda(s,m_b^2,m_c^2)}} \left\{\log^2(1-\omega)-\log^2(1+\omega) +2\log\frac{4s}{\bar{s}}\log\left(\frac{1+\omega}{1-\omega}\right)
+2{\rm Li_2}\left(\frac{1-\omega}{2}\right)\right. \nonumber \\
&&-2{\rm Li_2}\left(\frac{1+\omega}{2}\right)+2{\rm Li_2}\left(\frac{1+\omega}{1+M}\right)+2{\rm Li_2}\left(\frac{1+\omega}{1-M}\right)
-2{\rm Li_2}\left(\frac{1-\omega}{1-M}\right)-2{\rm Li_2}\left(\frac{1-\omega}{1+M}\right) \, \nonumber \\
&&\left. +\frac{2\omega \bar{s}}{s} -\frac{\bar{s}}{s}(1+\omega^2)\log\left(\frac{1+\omega}{1-\omega}\right) \right\} \, ,
\end{eqnarray}
where $\bar{s}=s-(m_b-m_c)^2$,
\begin{eqnarray}
\omega_1&=&\frac{\sqrt{\lambda(s,m_b^2,m_c^2)}}{s+m_b^2-m_c^2} \, ,\nonumber\\
\omega_2&=&\frac{\sqrt{\lambda(s,m_b^2,m_c^2)}}{s+m_c^2-m_b^2} \, ,\nonumber\\
M&=&\frac{m_b+m_c}{m_b-m_c}\, ,\nonumber\\
{\rm Li_2}(x)&=&-\int_0^x dt \frac{\log(1-t)}{t} \, . \nonumber
\end{eqnarray}

\end{document}